# A Case for Electron-Astrophysics

WHITE PAPER FOR THE VOYAGE 2050 LONG-TERM PLAN
IN THE ESA SCIENCE PROGRAMME

Contact Scientist: Daniel Verscharen

Mullard Space Science Laboratory
Department of Space and Climate Physics
University College London

Holmbury St Mary
Dorking
RH5 6NT

United Kingdom

E-Mail: d.verscharen@ucl.ac.uk
Telephone: +44 1483 20-4951

Voyage 2050 White Paper: A Case for Electron-Astrophysics

# 1  Core Proposing Team

**Contact Scientist:** Daniel Verscharen (University College London, United Kingdom)
**Deputy:** Robert T. Wicks (University College London, United Kingdom)

Olga Alexandrova (Observatoire de Paris, France)
Roberto Bruno (INAF, Italy)
David Burgess (Queen Mary University of London, United Kingdom)
Christopher H. K. Chen (Queen Mary University of London, United Kingdom)
Raffaella D'Amicis (INAF, Italy)
Johan De Keyser (BIRA-IASB, Belgium)
Thierry Dudok de Wit (LPC2E, France)
Luca Franci (Queen Mary University of London, United Kingdom)
Jiansen He (Peking University, China)
Pierre Henri (LPC2E, CNRS, France)
Satoshi Kasahara (University of Tokyo, Japan)
Yuri Khotyaintsev (Institutet för Rymdfysik, Sweden)
Kristopher G. Klein (University of Arizona, United States)
Benoit Lavraud (Institut de Recherche en Astrophysique et Planétologie, France)
Bennett A. Maruca (University of Delaware, United States)
Milan Maksimovic (Observatoire de Paris, France)
Ferdinand Plaschke (Institute for Space Research, Austria)
Stefaan Poedts (KU Leuven, Belgium)
Christopher S. Reynolds (University of Cambridge, United Kingdom)
Owen Roberts (Institute for Space Research, Austria)
Fouad Sahraoui (Laboratoire de Physique des Plasmas, France)
Shinji Saito (Nagoya University, Japan)
Chadi S. Salem (Space Sciences Laboratory, UC Berkeley, United States)
Joachim Saur (University of Cologne, Germany)
Sergio Servidio (University of Calabria, Italy)
Julia E. Stawarz (Imperial College London, United Kingdom)
Štěpán Štverák (Czech Academy of Sciences, Czech Republic)
Daniel Told (Max Planck Institute for Plasma Physics, Germany)

## Table of Contents

Cover page: Galaxy cluster MACS J00254.4-1222 from HST ACS/WFC and the Chandra X-Ray Observatory (Credit: NASA, ESA, CXC, UC Santa Barbara, Stanford University).

## 2 Electron-astrophysics processes and science questions

A grand-challenge problem at the forefront of physics is to understand how energy is transported and transformed in plasmas. This fundamental research priority encapsulates the conversion of *plasma-flow and electromagnetic energies into particle energy*, either as heat or some other form of energisation. The smallest characteristic scales, at which *electron dynamics* determines the plasma behaviour, are the next frontier in space and astrophysical plasma research. The analysis of astrophysical processes at these scales lies at the heart of the field of *electron-astrophysics*. Electron scales are the ultimate bottleneck for dissipation of plasma turbulence, which is a fundamental process not understood in the electron-kinetic regime. Since electrons are the most numerous and most mobile plasma species in fully ionised plasmas and are strongly guided by the magnetic field, their thermal properties couple very efficiently to *global plasma dynamics and thermodynamics*.

> Electrons determine the physics at the smallest characteristic scales in plasmas. The field of electron-astrophysics studies processes at these smallest scales in astrophysical plasmas. By utilising the solar wind as the prime and only accessible example for an unbounded astrophysical plasma, we propose to study electron-astrophysics through in-situ plasma measurements at electron scales.
>
> The key science questions of electron-astrophysics are:
>
> Q1. What is the nature of waves and fluctuations at electron scales in astrophysical plasmas?
>
> Q2. How are electrons heated and accelerated in astrophysical plasmas?
>
> Q3. What processes determine electron heat conduction in astrophysical plasmas?
>
> Q4. What is the role of electrons in plasma structures and magnetic reconnection?

The answers to these questions are fundamental to our understanding of the dynamics and thermodynamics of plasmas throughout the Universe: from the solar wind to stellar coronae, accretion discs, the intra-cluster medium, and even laboratory plasmas.

A plasma is an ionised gas in which mobile *ions* and *electrons* interact self-consistently and collectively with electromagnetic fields. Plasma is by far the most abundant state of baryonic matter in the Universe. Astrophysical plasmas, in general, exhibit a property called *quasi-neutrality*, which means that the total number of all ion charges is equal to the total number of electrons on global scales. As some ions are multiply charged (e.g., $He^{2+}$), electrons are the *most abundant particle species* in fully ionised plasmas. Nevertheless, an electron is 1836 times less massive than a proton, the lightest ion. Thus, ions typically dominate the *momentum flux*, but electrons and their associated kinetic processes dominate the *electrical and thermal conductivities*, making them hugely important for the plasma thermodynamics. Only recently, the field of plasma astrophysics has realised the importance of electron-scale physics for the evolution of the largest structures in the Universe and that limiting investigations to ion-scale physics would not solve the plasma-heating problem in the Universe.

### 2.1 Kinetic processes in electron-astrophysics

Almost all characteristic spatial and temporal scales associated with electron plasma physics are much smaller and shorter than the spatial and temporal scales associated with ion physics [e.g., in the solar wind, electron scales are of order a few 100 m, while ion scales are of order 100 km; 1]. The electron scales include the *electron gyro-radius* $\rho_e$, at which the electrons' gyro-motion about the magnetic field occurs, the *electron inertial length* $d_e$, at which the electron trajectories de-magnetise, and the electron Debye length $\lambda_e$, above which single-particle effects are shielded by neighbouring charges from the surrounding plasma.

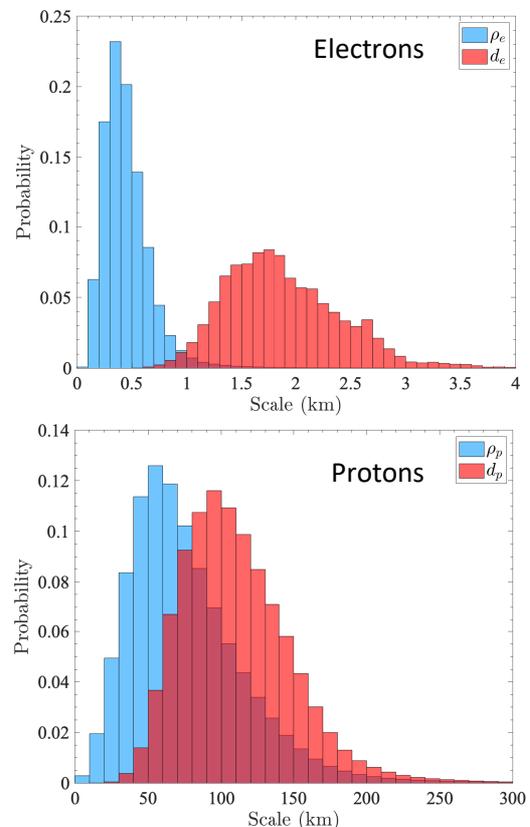

Figure 1: Probability distribution of characteristic electron scales (top) and proton scales (bottom) in the solar wind at 1 au from the Wind spacecraft.

**Table 1: Electron-astrophysics processes and the most important related unanswered science questions.**

| Type of interaction and associated electron distribution | Characteristics and observables | Key open science questions |
|---|---|---|
| **(A) Collisional relaxation (2.1.1)** 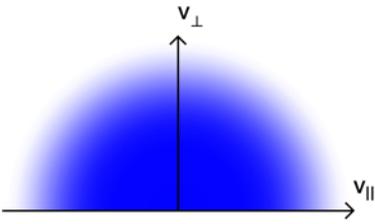 | • Quasi-isotropic, Maxwellian (core) distribution<br>• Reduced non-equilibrium features<br>• Equilibrates proton and electron temperatures<br>• Requires sufficiently high collision frequency | • Can collisions explain relaxation of non-thermal features and energy partition between ions and electrons?<br>• To what degree do collisions regulate electron heat flux?<br>• How do collisions dissipate the distribution's fine structure? |
| **(B) Expansion (2.1.2)** 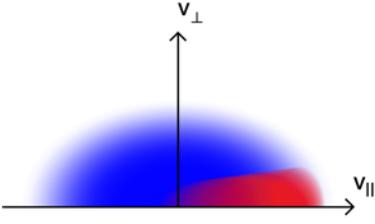 | • Double-adiabatic expansion creates temperature anisotropy<br>• Energetic electrons focus into strahl (red)<br>• Field-parallel energetic component with narrow pitch-angle spread | • How does the strahl form, and how does it vary with plasma conditions?<br>• What are the different effects of expansion perpendicular to or parallel to the magnetic field? |
| **(C) Instabilities (2.1.2)** 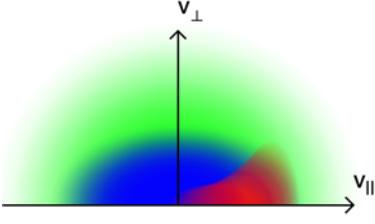 | • Anisotropy-driven instabilities reduce temperature anisotropy<br>• Strahl-driven instabilities broaden and scatter strahl into halo (green)<br>• Instabilities create electron-scale fluctuations | • How do fluctuations created by instabilities contribute to small-scale turbulence?<br>• How do instabilities regulate electron anisotropies, drifts and heat flux? |
| **(D) Landau damping (2.1.4)** 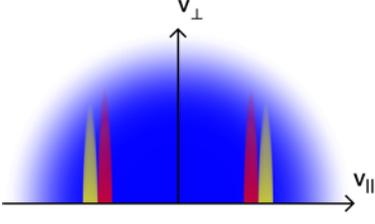 | • Produces enhancements (yellow) and depletions (red) around parallel phase speed, $v_\parallel = \omega/k_\parallel$<br>• Requires *fluctuations in parallel E-field* or *in parallel B-field (transit-time damping)*<br>• Parallel electron heating | • How important is Landau damping for electron heating?<br>• What fluctuations capable of Landau damping are active?<br>• How does Landau damping depend on plasma parameters? |
| **(E) Cyclotron damping (2.1.4)** 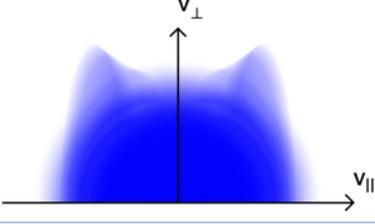 | • Produces shell-like extensions near resonance speed(s)<br>• Requires *fluctuations in perpendicular E-field*<br>• Perpendicular electron heating | • How important is cyclotron damping for electron heating?<br>• What fluctuations capable of cyclotron damping are active?<br>• How does cyclotron damping depend on plasma parameters? |
| **(F) Stochastic heating (2.1.4)** 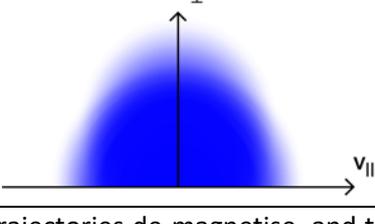 | • Produces diffusion of particles to larger $v_\perp$<br>• Requires *low-frequency but small-scale E-field fluctuations*<br>• Perpendicular electron heating | • How important is stochastic heating for electron heating?<br>• What fluctuations are most relevant for stochastic heating?<br>• How does this depend on plasma parameters? |

Figure 1 shows probability distributions of the electron and ion scales in the solar wind at 1 au. The global scale of this plasma is of order 1 au $\approx 1.5 \times 10^{11}$ m, while electron scales are just a few hundred metres. This *small-scale nature* of electron processes creates major challenges for their measurement and thus has stymied our attempts to understand electron physics. Likewise, electron-astrophysics processes occur on *timescales* that are much shorter than the timescales associated with ion processes. In the solar wind at 1 au, for instance, the electron gyro-period is of order tenths of a millisecond, while the ion gyro-period is of order seconds. *The study of electron-astrophysics aims to resolve these challenges to enhance our understanding of plasma physics throughout the Universe.*

*Space plasmas* (i.e., those in our solar system) are the only astrophysical plasmas in which spacecraft have measured plasma and field properties *in situ*. Solar-wind measurements indicate that electrons in astrophysical plasmas are typically not in *local thermodynamic equilibrium* [2-4]. Consequently, understanding how energy is transported, transferred, and dissipated through plasma electrons requires a very detailed analysis of the *electron velocity distribution function,* which fully describes the electrons' *kinetic behaviour*. Under the assumption that the distribution function is gyrotropic (i.e., cylindrically symmetric about the local magnetic field), it can be reduced to the two-dimensional gyrotropic distribution function in cylindrical coordinates $(v_\perp, v_\parallel)$ with respect to the local magnetic field. This two-dimensional function, which can be measured much more quickly than a full three-dimensional distribution function, is also called a *pitch-angle distribution* when transformed to energy and pitch-angle space. If *non-gyrotropic effects* are negligible, the pitch-angle distribution describes the properties of the plasma electrons completely [5]. Table 1 summarises the most important kinetic processes in electron-astrophysics and illustrates their characteristic signatures in the electron distribution. In the following, we discuss these different pathways for energy conversion.

### 2.1.1 Coulomb collisions

*Coulomb collisions*, soft scatterings between charged particles, relax deviations from thermal equilibrium and eventually dissipate fine structure in the distribution function, increase entropy, and heat the plasma. If collisions are sufficiently strong, the velocity distribution is *Maxwellian* (see Table 1A). However, decades of *in-situ* measurements of the solar wind have revealed that plasma electrons generally exhibit a complicated (i.e., *non-Maxwellian*) kinetic behaviour with fine structure that is consistent with partial but not total collisional relaxation [6-15]. Therefore, solar-wind electrons are affected by both *collisional* and *collisionless kinetic processes*. In a collisional system, the collisional timescales are much shorter than those associated with collective plasma processes. In a collisionless system, the collisional timescales are much longer than the timescales for collective processes. However, when fine structure is present in the distribution, collisional time scales become much shorter, even in very low-density plasmas [16].

### 2.1.2 Plasma expansion and kinetic instabilities

The observed non-Maxwellian features directly result from plasma expansion/compression, instabilities, and local heating, all of which are closely linked to the electromagnetic fields. *Expansion* drives temperature anisotropies due to double-adiabatic effects [17] and focuses energetic electrons into field-aligned beams [called "strahl" in the solar wind; 18; see also Figure 4 and Section 2.2] due to decreasing magnetic field strength (see Table 1B). Free energy in these non-equilibrium features in the electron distribution can drive *kinetic plasma instabilities*. These electron-driven instabilities reduce the free energy by modifying the distribution's shape through the creation of electromagnetic fluctuations at electron scales and subsequent particle scattering. These growing fluctuations combine with electron-scale fluctuations from the turbulent cascade to modify the overall thermodynamics and behaviour of the plasma [19-21]. Linear Vlasov-Maxwell theory reveals multiple potential sources of free energy in electron distributions to drive instabilities. Sufficient *electron temperature anisotropy*, for example, drives electron-scale instabilities [22-28]. *Electron heat flux* serves as another potential source of free energy for instabilities [29-32]. At high frequencies, electrostatic instabilities arise from the relative drift of different electron populations; e.g., anti-sunward strahl or counter-streaming strahls in the solar wind [33], and superthermal electrons provide additional free energy [34-36]. All of these instabilities create characteristic observable structures in the electron velocity distribution as they saturate.

Solar-wind strahl electrons are quasi-continuously transferred into another superthermal component called the "halo", which is isotropic and reaches energies above 100 eV [see Table 1C; 4, 37, 38]. Neither collisional effects nor the strahl broadening due to wave scattering [39] can fully explain this behaviour or the existence of the halo. Resolving this puzzle would be a major breakthrough in our understanding of superthermal, heat-flux carrying electrons in all collisionless plasma flows. This modification of electron heat flux by local small-scale instabilities [40-47] is one of many examples that show the necessity to understand electron-scale kinetic physics for the understanding of astrophysical flows.

### 2.1.3 Small-scale plasma turbulence

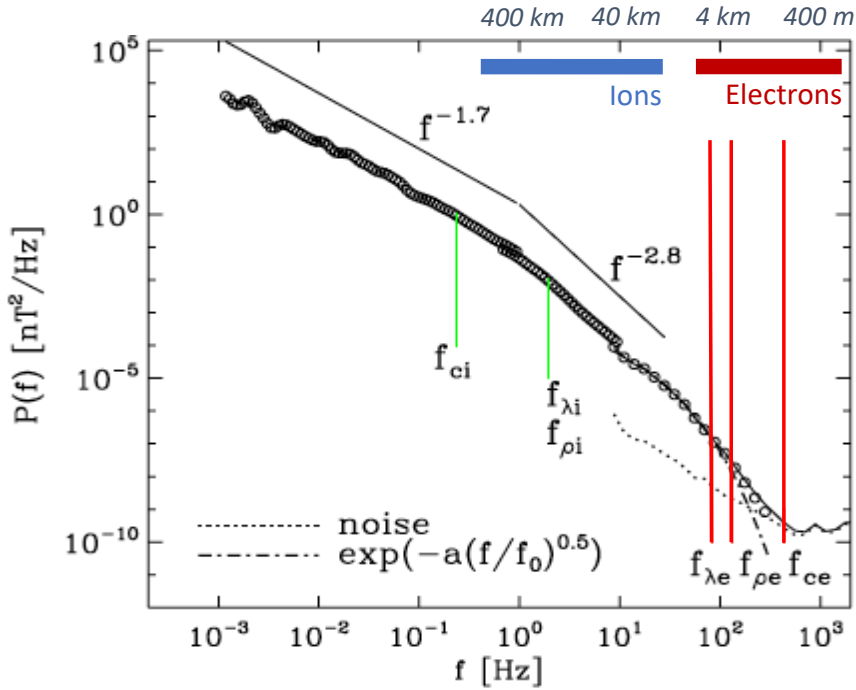

Figure 2: A turbulent power spectrum of the magnetic field computed using Cluster data. It ranges from fluid scales through ion scales to electron scales [51]. The coloured bars indicate the typical ion and electron scales. Previous missions have been capable of resolving electron scales only *temporally*.

Past measurements suggest that *local heating processes* of both ions and electrons have a substantial impact on the global thermodynamics of space plasmas. For example, radial profiles of the solar wind's temperature reveal a much slower cooling rate than expected for an adiabatically expanding gas [48-50]. The *dissipation of plasma turbulence* is considered the leading paradigm for the heating of particles in plasmas. *In-situ* observations have shown that plasma turbulence develops a cascade that transports energy from large-scale flows and fields down to small, kinetic scales at which the energy dissipates and heats the particles. This *energy cascade* is apparent in power spectra of the magnetic-field fluctuations (e.g., Figure 2). *Fluctuations at large scales* (greater than a few hundred km), at which magnetohydrodynamic (MHD) theory is applicable, have been studied for decades, and – although some aspects remain uncertain – a consistent picture of their behaviour has emerged [52-55]. Conversely, *fluctuations at small scales*, at which particle heating and dissipation occur, are governed not by MHD theory but by a complex interplay of poorly understood kinetic mechanisms. In this *kinetic range*, on scales comparable to the ion and electron gyro-radii ($f_{\rho i}$ and $f_{\rho e}$ in Figure 2), we expect:

- wave modes become *dispersive* and alter their character;
- collisionless *field-particle interactions* transfer energy between fields and particles, either by:
  - damping electromagnetic fluctuations and energising particles, or conversely
  - exciting field fluctuations through *kinetic instabilities;* and
- dissipative *coherent structures*, such as *current sheets* or *vortices*, form.

The relative contributions of these mechanisms to plasma electron heating currently remain unknown, though these mechanisms are universal and important in all astrophysical plasmas.

As the turbulent cascade approaches electron scales, *electrostatic modes* become increasingly important rather than the electromagnetic modes that dominate fluctuations at larger scales. We distinguish between two types of electrostatic fluctuations [56-58]. First, low-level spontaneous *quasi-thermal noise emissions* are present – even in the absence of free energy to drive instabilities – as random emissions of the plasma particles. Second, *induced electrostatic fluctuations* with higher amplitudes are either locally generated by *kinetic instabilities* or result from a nonlinear decay of large-scale fluctuations triggered and convected by global plasma flows. Recent observations of the full spectrum of spontaneous electrostatic quasi-thermal noise fluctuations reveal peak intensities at high frequencies and in all directions of propagation with respect to the background magnetic field [59, 60]. Although electron energisation through electrostatic modes is a universal process in small-scale plasma turbulence – from laboratory (e.g., electrostatic gradient-driven turbulence) to astrophysical (e.g., beam-generated turbulence in stellar flares) plasmas, the quantitative polarisation, anisotropy, and nonlinear properties of these energy channels remain unknown. Therefore, future electron-astrophysics measurements must allow us to quantify the amplitude, frequency, and occurrence rate of electrostatic modes. Measuring electrostatic fluctuations at the plasma frequency will also provide an independent and fast measurement of the electron density for cross-calibration [61-65].

### 2.1.4 Dissipation at electron scales

In all weakly collisional plasmas, heating is a two-step process. First, collisionless interactions reversibly transfer energy to the particles. Then, the distribution function develops fine structure in velocity space which raises the efficiency of collisions [even though the plasma is overall still weakly collisional; 16]. The collisions then irreversibly thermalise the energy in the particles and heat the plasma [66-69]. *Field-particle interactions* governing the first step are classified as *resonant* vs. *non-resonant* interactions (Figure 3).

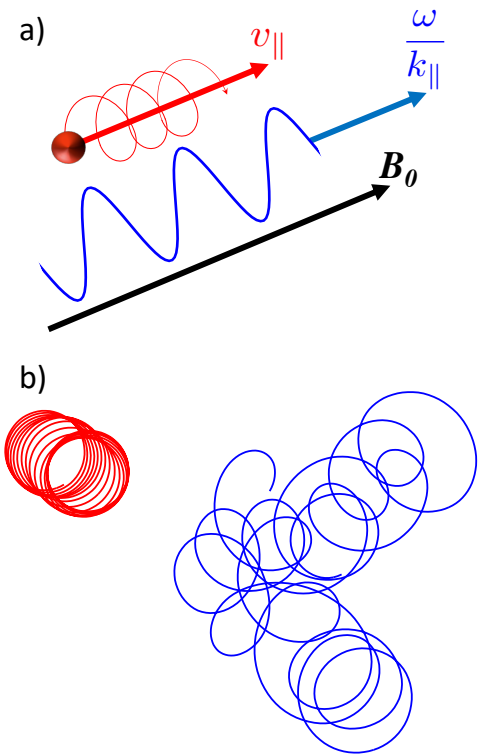

Figure 3: Electron dissipation mechanisms. a) A Landau-resonant electron in a monochromatic wave. b) Particle orbits in small-amplitude (red) and large-amplitude (blue) gyro-scale fluctuations, leading to stochastic heating.

Resonant interactions include *Landau damping*, *transit-time damping*, and *cyclotron damping* (Figure 3a). In Landau damping, for example, if an electron's velocity component $v_\parallel$ parallel to the magnetic field matches the parallel phase speed $\omega/k_\parallel$ of a wave, it resonates with the wave electric field, leading to energy transfer from the wave to the particle. Such *collisionless damping mechanisms* re-shape the distribution function and create the characteristic signatures shown in Table 1D&E. They are well understood for individual waves; however, we are only beginning to understand them in the nonlinear regime of *strong turbulence*.

Non-resonant interactions include *stochastic heating* [70, 71] and *magnetic pumping* [72, 73]. We illustrate stochastic heating in Figure 3b. If the amplitude of electric or magnetic fluctuations on the spatial scales of the electron gyro-motion is small (left-hand side in red), the particle's orbit is circular and drifts due to the large-scale changes in the field. Conversely, if the amplitude of the gyro-scale fluctuations is large (right-hand side in blue), the orbits are perturbed and become stochastic. The acceleration due to the fluctuating electric fields then leads to a diffusion in kinetic energy in the direction perpendicular to the magnetic field [74] and thus an extension of the distribution to greater perpendicular velocities (see Table 1F). Due to the lack of appropriate measurements, these dissipation mechanisms have never been compared at electron scales in a turbulent astrophysical plasma. *Understanding their relative importance will achieve breakthroughs in our interpretation of observations and our modelling capabilities of electron thermalisation.*

In a further complication, heating occurs *intermittently* in fluid and plasma turbulence, i.e., in spatial and temporal bursts [75-82]. *Intermittent structures*, such as short-lived small-scale current sheets, can harbour localised electron energisation through *magnetic reconnection* which also efficiently feeds the turbulence spectrum at electron scales through rapid current-sheet formation and disruption [83-85]. Intermittent energisation may also occur at *shocks* or *double layers* [86, 87]. These structures are associated with wave emission [88, 89], which can in turn heat particles. Collisional effects are also mostly concentrated in the proximity of these structures [90], which are of order the characteristic electron scales. The analysis of these structures thus requires electron measurements to quantify the associated energy transfer and the resultant features in the electron distribution on small scales. *In order to make ground-breaking observations of coherent structures in small-scale turbulence, we must resolve electron distribution functions within small structures.*

> Our approach to answer fundamental electron-astrophysics questions through measurements in the solar wind requires us to disentangle collisional, expansion, instability, and dissipation effects in the solar wind in order to resolve our key science questions. We must understand the relative importance of these processes in the solar wind and extrapolate our results to other astrophysical plasmas.

## 2.2 Electron-astrophysics in the solar wind

Previous measurements of particle properties show that solar-wind electrons undergo both collisional and collisionless kinetic processes. Electrons have undergone between 0.1 to 1000 collisions on their way from the Sun to 1 au, making 1-au solar wind the ideal testbed to study both classes of interactions in the astrophysical context [91]. It is the only unbounded astrophysical plasma accessible to *in-situ* measurements, and even laboratory plasmas cannot be measured to the same degree of accuracy. In the solar wind (and presumably in all plasma outflows), electrons create an ambipolar electric field that contributes to the acceleration of the plasma flow via thermal-pressure gradients and an ever-present tail in their distribution [92-97]. However, the details of this *exospheric* contribution to the overall plasma dynamics remain unknown.

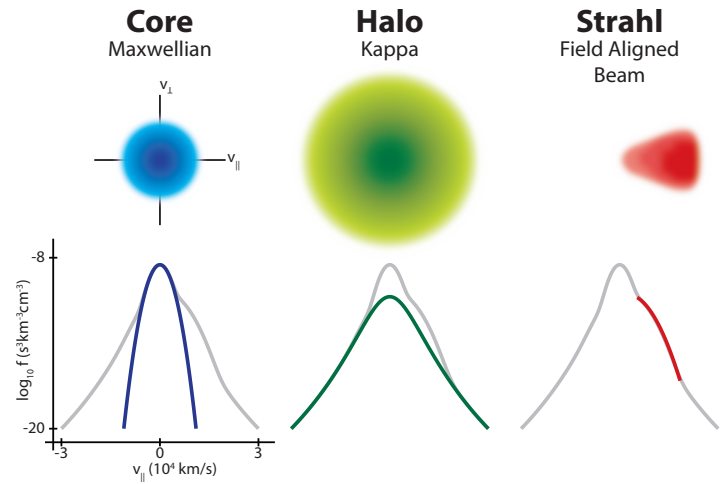

Figure 4: Typical components of solar-wind electron distribution functions in velocity space: core, halo, and strahl. Credit: M. Pulupa.

The solar-wind electron components core, halo and strahl [see Figure 4; Table 1B; 7, 10, 98-102] at times exhibit temperature anisotropies [9, 27]. The core typically includes about 95% of the electrons. While halo electrons can travel in all directions, the strahl appears as a highly focused, magnetic-field-aligned beam that moves predominantly away from the Sun [11]. The strahl also serves as a tracer for changes in the magnetic-field topology [103, 104]. Although a number of models exist that explain the halo formation through turbulent electron acceleration, quasi-thermal noise, or the interaction with instability-driven waves [105-108], past observations are insufficient to distinguish among them. The multi-component structure of the electron distribution carries a significant *electron heat flux* [109, 110]. Understanding heat-flux regulation is critical for the development of global models for the solar wind and other astrophysical plasma flows, but the relative importance of the relevant mechanisms remains unclear.

The solar wind is an excellent medium to study astrophysical plasma turbulence and turbulent heating under varying plasma conditions. Measurements in the fast solar wind [111-115] and numerical simulations [116-124] suggest that the nature of the heating mechanism depends on species, plasma conditions, and potentially the different physics at the source regions of the plasma flow. Conversely, measurements of the slow solar wind reveal electrons to be hotter than protons [91]. Previous multi-point observations at ion scales [125, 126] and sub-ion scales [127, 128] also show that the nature and occurrence of *intermittent structures* differ between slow and fast solar wind. Slow wind exhibits a greater variety of such structures – compressible vortices, solitons, and shocks – which reflects the plasma's origin in the regions of closed magnetic fields in the corona. Conversely, fast wind is typically less complex, containing fewer compressive features. This heterogeneity in plasma parameters demonstrates again that the solar wind at 1 au is an ideal plasma laboratory and provides all *in-situ* spacecraft with a broad variation in plasma conditions to be sampled. Although we cannot observe the full expansion of the solar wind by measuring its *in-situ* properties at 1 au, this kind of measurement allows us to observe the plasma processes that cascade energy to smaller scales and the energisation at these scales.

> The space-plasma community's experience shows that the solar wind provides a unique means for observing kinetic processes and turbulence. *We propose to exploit this fact to answer the open questions of electron-astrophysics. This field of research will achieve major breakthroughs in our understanding of the Universe.*

## 2.3 Electron-astrophysics outside the solar system

Detailed observations of the solar wind provide insights into the plasma processes in more remote systems. Research in the field of electron-astrophysics is the backbone for missions like Athena that will study x-ray emissions generated by heated and accelerated plasma electrons. We discuss two examples of contemporary astrophysical problems that will be substantially advanced by *in-situ* studies of electron-astrophysics.

### 2.3.1 Heat transport in the intracluster medium of galaxy clusters

*Galaxy clusters*, some of the largest gravitationally bound structures in the Universe, have an interesting architecture: most of the mass resides within a large (>3-million light-year radius), approximately spherical distribution of dark matter. Most of the cluster's baryons, however, reside in the hot ($10^7$ < T < $10^8$ K) and tenuous ($10^{-3}$ < $n_e$ < $10^{-1}$ cm$^{-3}$) *intracluster medium (ICM)*, which is in hydrostatic equilibrium with the gravitational potential of the dark matter [129]. Figure 5 shows an x-ray image of a galaxy cluster superposed on an optical image. The galaxies themselves comprise a tiny fraction of the cluster's mass and act, in this sense, as tracers. Based on x-ray observations, the *radiative cooling time* in the central "core" of the ICM would be about $10^8$ yr, which is short compared to typical cluster lifetimes. At this rate, each year, 100's to 1000's of solar masses of gas would cool into molecular clouds, which would result in highly-active *star formation* in the cluster's core. Nevertheless, no such phenomenon is observed: while ICM cores host some cold gas, the amount is far too low to support prodigious star formation. This apparent contradiction can only be reconciled by the action of some ongoing *heating processes* on the core. Candidate mechanisms include (1) *inward heat transport* from outer regions, or (2) *active galactic nucleus (AGN) feedback*: heating by jets of supermassive black holes. Resolving this problem of ICM dynamics is critical for understanding the formation and evolution of the most massive galaxies. *Thermal conduction* may play a critical role in ICM heating. First, it can enable direct flow of heat into the core from the hotter outer regions, thereby reducing the required energy injection from the central black hole [130]. Electrons dominate the thermal conduction due to their large mobility. Second, they contribute to AGN feedback by dissipating acoustic waves generated by jets and regulating local instabilities that enable flows of cold gas to feed the AGNs through precipitation.

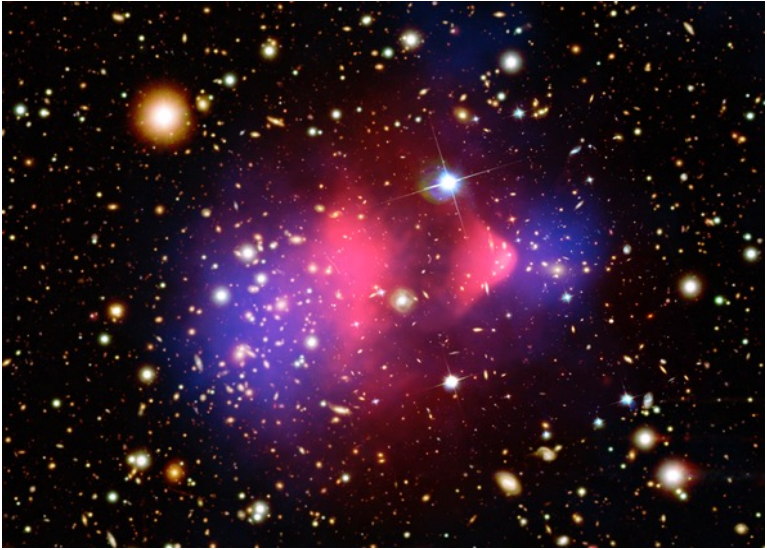

Figure 5: X-ray image of the Bullet cluster (Credit: x-ray: NASA/CXC/CfA/M. Markevitch et al.; optical: NASA/STScI, Magellan/U. Arizona/D. Clowe et al.; lensing map: NASA/STScI ESO WFI, Magellan/U. Arizona/D. Clowe et al.).

The physics of heat transport in a weakly collisional, high-$\beta$ (>10) plasma, where $\beta$ is the ratio between thermal and magnetic pressure, including the ICM, remains poorly understood. Recent simulations highlight the role of heat-flux-driven whistler modes in modulating the *electron heat flux*, suggesting that heat is transported at essentially the whistler phase speed [131]. The resulting dependence of heat flux on plasma properties fundamentally differs from that in collisional *fluid* theory. This has macroscopic implications for the ICM by, e.g., changing the conditions for *local thermal instability*, which facilitates the fuelling of AGNs. However, the current theoretical models are based on particle-in-cell simulations that can only achieve a factor of 100 separation between the electron gyro-radius and the temperature scale length. These scales are separated by a factor of $10^{12}$ in the real ICM, rendering the inferred impact of these kinetic phenomena on ICM processes highly dependent upon the analytical extrapolation of the models (e.g., quasi-linear vs. particle trapping).

Electron-astrophysics missions must generate high-frequency and multi-point measurements of magnetic-field fluctuations and high-cadence measurements of electron distributions that will further our understanding of the degree to which *electron heat transport* across astrophysically relevant scales is possible under different circumstances [132]. This will allow us to test the applicability of *quasilinear theory* and/or *particle-trapping models* to whistler-mediated thermal conduction [131, 133] in a setting where there is a large scale separation ($\sim 10^6$) between the electron gyro-radius and the temperature scale length. *These results will enable major improvements in models for the heating of the ICM and similar astrophysical environments.*

### 2.3.2    Ion vs. electron heating in accretion discs

In many astrophysical plasmas, collisions between ions and electrons are extremely infrequent compared to dynamical processes and to collisions within each species. It is an important open question whether, in the effective absence of interspecies collisions, there is any mechanism for the system to self-organise into a state of equilibrium between the two species and, if not, what sets the *ion-to-electron temperature ratio*. This question of fundamental plasma physics also carries particular importance for understanding astrophysical regions of *radiatively inefficient accretion flows* onto black holes such as the one at our own Galactic Centre, Sgr A*. Two basic scenarios have been theorised to account for the observed low luminosity of such accretion discs [134]: (1) The proton *heating rate* (e.g., through the dissipation of plasma turbulence) exceeds that of the electrons ($Q_p/Q_e \gg 1$). As a result, most of the thermal energy is imparted to ions which do not efficiently radiate before entering the black hole. (2) The protons and electrons have similar heating rates ($Q_p/Q_e \sim 1$), but the accretion rate is low. As a result, most of the plasma is carried away by *outflows* rather than entering the black hole. Global magnetohydrodynamic models that seek to distinguish these two theories rely heavily on accurate heating prescriptions: theoretical or observational formulae for $Q_p/Q_e$ as functions of $\beta$ and the proton-to-electron temperature ratio $T_p/T_e$. Resolving the relative heating of protons and electrons has important implications since the proton-to-electron relative heating rate directly affects the accretion rate and the formation of outflow jets [135, 136]. Interest in this science question is being fuelled by the advent of the *Event Horizon Telescope* [137], which published the first picture of a black hole surrounded by a low-luminosity accretion disc [see Figure 6; 138]. The radio emission seen in these measurements results from heated electrons gyrating in the magnetic field of the inner accretion disc. Both the appearance and the gross dynamics of the inner accretion disc are crucially dependent upon these (still uncertain) heating models. For example, global simulations of accretion discs with low ion heating have found a radiating jet but no visible jet with a more equitable heating model.

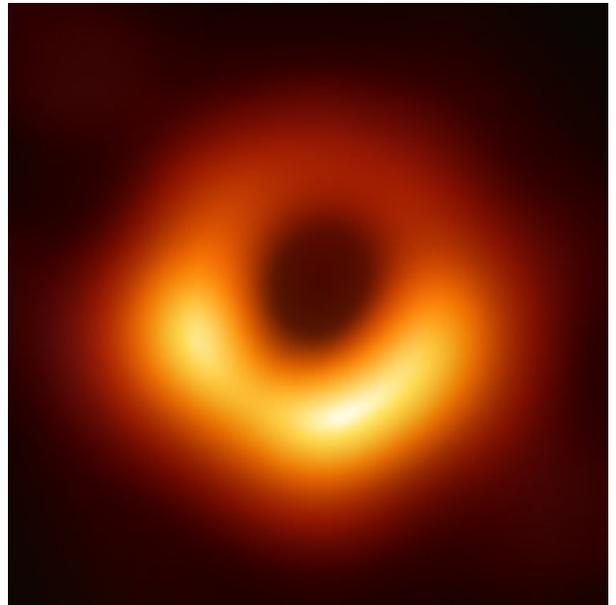

Figure 6: First image of a black hole (M87*) from the Event Horizon Telescope [138]. All of the 'light' (i.e., radio waves) seen in this image is created by heated and accelerated plasma electrons in the accretion disc's magnetic field.

Electron-astrophysics missions must make measurements of electron-scale plasma turbulence that will enable us to probe two key aspects of this problem: the *nature of the fluctuations* (Q1) and the *amount of heating* (Q2). Under certain assumptions about the nature of the turbulence (e.g., that it consists of low-frequency, anisotropic perturbations of the kinetic-Alfvén type), it is possible to prove that any turbulent cascade of such fluctuations found at sub-ion scales is destined for electron heating [139]. We must thus measure the fraction of the turbulent energy going into electrons as a function of ambient plasma parameters, viz., $\beta$ and $T_p/T_e$. Moreover, via these fluctuation measurements, we must determine whether the underlying assumptions mentioned above are indeed true and so whether there is, in fact, a significant part of the turbulent energy that is channelled into ions via cyclotron heating or stochastic heating due to deformations of ion Larmor orbits [Section 2.1.4; 71]. Measurements of the perturbations in the proton distribution at sub-Larmor scales are also required to directly determine the amount of proton heating. Based on these studies, we will develop scaling relations for the heating rate that connect our measurements in the solar wind to the conditions in other astrophysical objects.

> Though we propose *in-situ* electron measurements in the solar wind, their results will be universal and as such also apply to other space plasmas such as the solar corona, the Earth's magnetosphere, and magnetospheres of other planets in or outside the solar system. More broadly, we consider the solar wind as representative of a myriad of astrophysical plasmas strewn throughout the Universe.

## 2.4 Electron physics in laboratory plasmas

Some plasma processes exhibit similar behaviour in space/astrophysical plasmas and laboratory plasmas. For example, scaling relations exist between astrophysical and laboratory environments [140, 141], underlining the complementarity in these two regimes. While such scaling relations are not always perfect (e.g., regarding dissipation coefficients from anomalous resistivity or viscosity, and due to limitations from edge effects), the similarity is sufficiently close to link astrophysical and laboratory plasmas for mutual benefit.

Turbulence driven by *electron temperature gradients (ETGs)* is a topic of particular interest in the laboratory-plasma community [see Figure 7; 142-144]. ETGs are thought to be the main drivers of anomalous *electron heat loss* in magnetically confined fusion plasmas. Such losses of heat and particles limit the confinement time and thereby constrain the feasibility of *fusion reactors* [145], which – so far – can only be overcome by increasing the reactor size (and hence, by substantially increasing the cost). Unfortunately, the high temperatures of fusion plasmas have largely limited observational studies of them to remote (versus *in-situ*) measurement techniques such as microwave reflectometry [146]. Although ETG turbulence has been successfully reproduced in *linear laboratory devices* [147-149], even this technique carries significant limitations due to the lack of access to measurements of particle distribution functions. While we do not expect the quiescent solar wind to exhibit ETG instabilities, we anticipate that these modes are excited during transient events, such as those reported by Roberts et al. [150]. In order to unambiguously identify ETG modes if they occur in the solar wind, we require simultaneous, *in-situ* measurements of electric fluctuations, magnetic fluctuations, and electron distributions. Such modes would result in a *heat-flux boost*, which significantly impacts the local heating of the plasma. While fusion devices usually operate at very low $\beta$ (<$10^{-3}$), the ETG instability has been theorised to couple to whistler waves [*W-ETG*; see 151, 152] at higher $\beta$ (~0.1). Observing if and how this transition occurs will provide a benchmark test of the plasma models that are currently applied to laboratory and astrophysical environments. In particular, the workhorse tool in the fusion community – *gyrokinetic models* – contains the ETG instability but not the whistler-wave mode. Advanced models containing both kinds of waves are currently under development and would benefit greatly from electron-scale observations of W-ETG coupling to fill this gap of understanding.

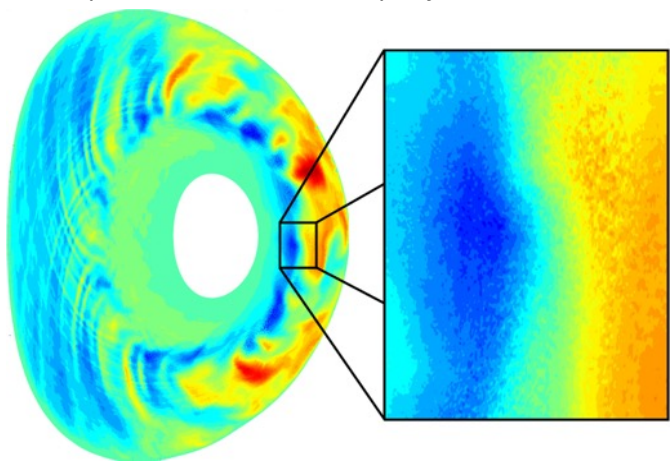

Figure 7: Simulation of ETG turbulence in the TCV-Tokamak (Credit: D. Told).

Laboratory experiments with *high-power lasers* offer a complementary approach to study *electrostatic* and *lower-hybrid turbulence* [153] for conditions that are relevant to cosmic plasmas, although they do not provide the same scale separations as astrophysical plasmas. For instance, these types of turbulence have been invoked to explain electron acceleration in *solar flares* and the kHz emission observed by the Voyager spacecraft near and beyond the *heliopause* [154, 155]. *An electron-astrophysics mission must study these types of electrostatic electron-scale fluctuations in order to understand their nature and their impact on electron thermodynamics.*

## 2.5 Science questions

Resolving the central challenges in the field of electron-astrophysics requires a programme of measurements on small plasma scales. We must identify the nature of the electron-scale fluctuations (Q1); characterise the dissipation and acceleration mechanisms at work (Q2); determine the processes that determine electron heat conduction (Q3); and investigate the role of electrons in plasma structures and reconnection (Q4). This electron-astrophysics research programme tackles the key problem of *understanding the behaviour of energy in the Universe*. It will help us understand the heating mechanisms the are responsible for the creation of UV and x-ray emissions observed throughout the Universe.

### 2.5.1 Q1. What is the nature of waves and fluctuations at electron scales in astrophysical plasmas?

The plasma mechanisms that govern electron heating and acceleration depend critically on the nature of turbulent fluctuations at small, electron-kinetic scales (a few 100 m in the solar wind). Therefore, the first task in electron-astrophysics is to *identify the nature* of these small-scale fluctuations. The *critical-balance principle* [156-158] predicts that, in strong plasma turbulence, the nonlinear plasma response has a magnitude similar to the linear plasma response. This behaviour is consistent with solar-wind observations, including the predicted modifications to the field-fluctuation properties in the ion-dissipation range [159-162]. In this critical-balance paradigm, the identification of the nature of turbulent fluctuations is thus informed by the linear properties of the fluctuations – even in fully nonlinear plasma turbulence [163].

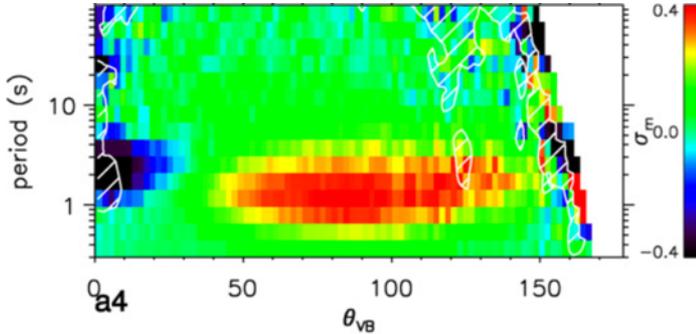

Figure 8: Magnetic helicity $\sigma_m$ at ion scales as a function of the angle between solar-wind flow and magnetic field $\theta_{VB}$ and fluctuation period. At small angles, negative values suggest ion-cyclotron / whistler waves. At large angles, positive values suggest kinetic-Alfvén turbulence [169].

*At large scales*, turbulence is predominantly non-compressive and shows correlations known from the *Alfvén plasma wave* [164]. Only a small fraction of energy is in compressive modes at large scales [165-167]. *At ion-kinetic scales*, the fluctuations transition to another regime [*kinetic-Alfvén turbulence*; 168]. Analyses of fluctuations at these scales reveal additional components generated by ion instabilities. For example, the method illustrated in Figure 8 reveals two components of ion-scale fluctuations: a narrow band (blue) with parallel wavevectors, likely to be waves driven by ion instabilities [169-174] and a broader band of kinetic-Alfvén waves (red) resulting from the turbulent cascade itself. Other ion-scale instabilities have also been identified [173, 175, 176].

*At electron-kinetic scales*, however, the nature of the turbulent fluctuations is not well understood. A variety of wave types can exist at these small scales: e.g., whistler waves, Bernstein waves, lower-hybrid waves, the recently predicted inertial kinetic-Alfvén waves [177-181], and electron-driven instabilities [e.g., 177]. Strong turbulence is known to generate intermittent coherent structures, such as *current sheets* [182-184], *electron-scale holes, vortices* [Figure 9; 185], *mirror modes*, *shocks*, and *double layers*. Electron-astrophysics missions must identify the nature of electron-scale fluctuations through high-cadence, multi-point electromagnetic-field measurements and subsequent *polarisation analysis* like the one shown in Figure 8. This knowledge will allow us to identify intermittent structures and to characterise the turbulence at this poorly understood *end of the turbulent cascade* where the electrons are energised as the turbulence fully dissipates.

We require measurements that enable the analysis of *power-law energy spectra* and the *anisotropic distribution of power* in wavevector space to help us explore the nonlinear evolution of electron-scale turbulence beyond the identification of its linear response. Previous missions, such as Cluster and MMS, have used multi-point measurements to determine these properties down to ion scales [160, 186], but many features of the turbulence at electron scales have not been measured due to the need for *high cadence, high sensitivity, and small spacecraft separation* to resolve this challenge at such small scales. Future missions must close this gap by simultaneously observing the electron-scale energy spectra of the electric and magnetic fields and making high-speed measurements of the electron distribution. This approach will enable us to discern the turbulent cascade and the wave generation through instabilities by analysis of peaks, breaks, and other spectral features. These simultaneous multi-point measurements must *disentangle spatial and temporal fluctuations*. Every separation between two measurement points samples one scale of spatial variation at a time. However, since turbulence continuously cascades across scales, we must measure fluctuations with multiple scale separations in order to understand the energy flow through wavevector space across scales. The field of electron-astrophysics thus requires *multi-scale missions* (either through measuring multiple scales sequentially or, ideally, through measuring them simultaneously). Multi-point measurements also provide the spatial structure of the fluctuations: e.g., reveal any elongation along or across the field and gauge gyrotropy about the field axis. Moreover, we must explore intermittency properties through statistical measures, such as structure functions, kurtosis, partial variance of increments [PVI; 187-190], local estimates of the turbulent energy transfer [191], and direct multi-point sampling, to determine their occurrence rate and their contribution to electron heating. These key turbulence properties establish the conditions under which the electron energisation operates, helping us to *constrain and identify the processes responsible for heating at the end of the turbulent cascade*.

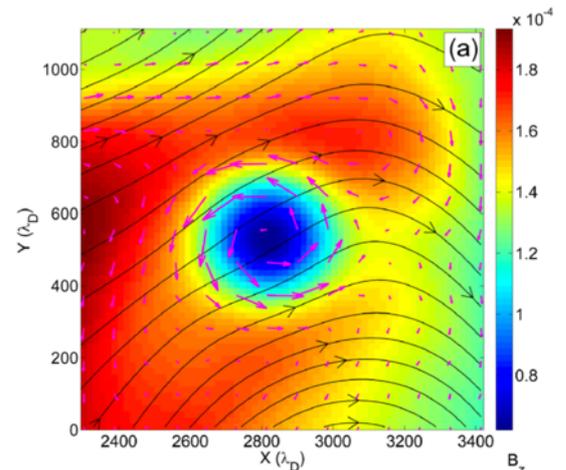

Figure 9: Simulation of a magnetic-hole structure: magnetic field (black), electron flow vectors (magenta), and parallel magnetic field (colour-coded) in a coherent electron-scale vortex [185].

### 2.5.2 Q2. How are electrons heated and accelerated in astrophysical plasmas?

The interaction between electrons and electromagnetic fields is the crucial link for electron heating. Future electron-astrophysics missions must enable the application of techniques such as the *field-particle correlation* method [192-194] to identify the acting heating mechanisms and compute the rate of electron energisation using single-point measurements. The method distinguishes the various energisation mechanisms listed in Table 1 and Section 2.1.4 by highlighting which regions in velocity space gain energy. Figure 10 shows the first application of the field-particle correlation technique using MMS measurements in the Earth's magnetosheath [194]. Here, the velocity-space signatures of *Landau damping* around the expected resonance speed are clearly visible in the correlation between the electric field and the particle distribution (shown in colour-coding). The rate of electron energisation is comparable to the estimated turbulent cascade rate, providing us with even more evidence that the dissipation of small-scale turbulence plays a critical role in electron heating. However, MMS lacks sufficient sensitivity for applying this method in the solar wind and lacks sufficient cadence to properly resolve heating at electron scales. Future electron-astrophysics missions must have *superior capabilities* – especially through high-cadence electron measurements – that will allow us to apply this technique at electron scales in the solar wind to understand the damping of turbulence, the action of instabilities, and other mechanisms leading to electron heating. These measurements require high time resolution to also enable the sampling of intermittent structures on electron scales by resolving spatial variation in the electron distribution on electron scales, which could not be achieved with previous missions.

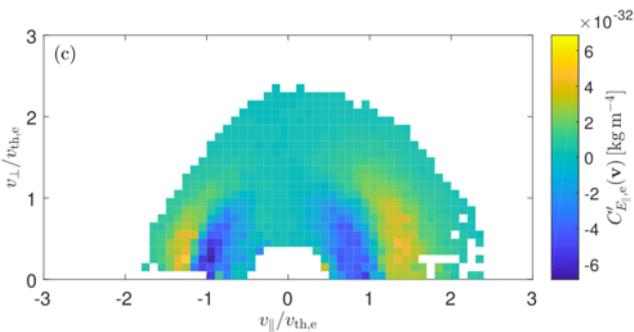

Figure 10: Energy transfer from turbulent magnetosheath fluctuations to electrons as a function of perpendicular and parallel electron velocity, measured by applying the *field-particle correlation technique* to MMS data with 30 ms cadence [194].

We must directly measure high-resolution electron velocity distributions organised by pitch angle in order to reveal the energy flow in velocity space and to determine how irreversible heating is achieved using such measurements. This entails measurements that enable us to examine the anisotropic *velocity-space cascade* by quantifying the fine structure of the particle distribution function [195, 196].

*Collisionless shock waves* are locations of strong particle acceleration [197, 198]. Turbulent electric fields in combination with the shock geometry can create conditions for shock drift acceleration or diffusive shock acceleration [199, 200]. Shock acceleration is an important plasma-energisation mechanism throughout the Universe reaching from cluster shocks [201] to supernova remnants [202], interplanetary space [203, 204], and stellar coronae [205]. In order to understand the electron-kinetic physics of shock acceleration to near-relativistic and relativistic energies, we require measurements of the energy spectra of *energetic electrons* up to multiple tens of MeV with simultaneous measurements of the shock properties. This includes measurements of the particles, fields, and the associated turbulence.

In order to understand the pathways to dissipation and acceleration, an electron-astrophysics mission also requires ion measurements to quantify the *partitioning of energy* between ions and electrons and the dependence of heating on different *plasma conditions*. These measurements include the proton temperatures and features in the proton distribution function (albeit at a suitably lower cadence corresponding to the ion scales that are generally larger than the electron scales) simultaneously with the rapid electron measurements in order to quantify the increase in internal energy in protons and electrons [206-211]. Resolving this issue is crucial for our understanding of the overall plasma thermodynamics because it directly quantifies the energy transfer in the system.

To bring complete closure to this science question, electron-astrophysics missions must cover *large statistical datasets* of the dominant electron-scale fluctuations and the mechanisms that transfer field energy into the plasma components depending on plasma conditions. This task must provide extrapolatable and quantitative results on the relevance of plasma modes and heating mechanisms in different astrophysical plasma environments and thus transform our understanding of the thermodynamics of plasmas throughout the Universe. These missions must sample different streams of the solar-wind plasma with a variety of background parameters over its mission lifetime. These parameters include the *ion-to-electron temperature ratio,* the *solar-wind bulk velocity,* and the *turbulence amplitude* [212, 213]. In addition, $\beta$ is one of the most critical plasma parameters [139, 214]. We must sample different *types of solar wind* that are comparable with other space or astrophysical plasmas. For example, *interplanetary coronal mass ejections* exhibit a low $\beta$ allowing us to probe plasma conditions similar to those in solar/stellar coronae and laboratory plasmas; while the fast solar wind can reach $\beta > 10$, allowing us to probe plasma conditions similar to accretion discs and the ICM. Even the damping rates of plasma modes are sensitive to parameters such as $\beta$ or the ion-to-electron temperature ratio, and thus their overall contribution to the energy budget depends critically on the background plasma parameters. In addition, electron-astrophysics missions must be capable of sampling a variety of non-thermal features in order to explore multiple plasma instabilities (see Section 2.1.2).

The *turbulence context* (e.g., compressibility and overall turbulence level) at multiple scales is another key property when measuring the dissipation of energy. For example, previous magnetosheath measurements show that the energy transfer rate at ion scales is enhanced at times of increased density fluctuations [215]. A similar measurement of the energy cascade rate at sub-ion scales has only recently been possible [216]. An electron-astrophysics mission will encounter broad variations in the turbulence context over its lifetime, which will facilitate a more complete picture of the evolution and behaviour of turbulence.

### 2.5.3　Q3. What processes determine electron heat conduction in astrophysical plasmas?

Due to their very high mobility, plasma electrons can carry large amounts of heat for long distances. The third velocity moment (skewness) of the electron distribution function at a given point in space characterises the heat flux carried by the electrons. If electron collisions are sufficiently frequent, the heat flux along the magnetic field follows the predictions by Spitzer and Härm [217], which assume a small deviation of the distribution function from the Maxwellian equilibrium. If electron collisions are very rare, however, the maximum available heat flux is given by the *free-streaming heat flux* under the assumption of a subsonic electron flow [218-220]. Recent observations of the solar-wind electron heat flux suggest that both the Spitzer-Härm and the free-streaming regimes can occur in the solar wind [see Figure 11; 221].

A large heat flux (i.e., a strong third velocity moment) represents a strong deviation in the electron distribution from its equilibrium state. If this deviation crosses the threshold of a *heat-flux driven kinetic micro-instability*, the plasma generates electromagnetic fluctuations on electron scales that scatter electrons in velocity space. Like in all kinetic micro-instabilities, this scattering mechanism reduces the source for instability which, in this case, is the heat flux itself [41, 43, 222]. These instabilities, once excited, thus limit the heat flux to a value below the free-streaming heat flux. At the same time, the unstable electromagnetic fluctuations at electron scales act as scattering centres for the electrons in configuration space [21]. The configuration-space scattering is similar to the action of binary Coulomb collisions and reduces the mean free path of the electrons. This self-regulation of heat flux by instabilities changes the overall *heat conduction and particle transport* in the plasma.

Like all kinetic processes, plasma heat-flux regulation is directly associated with structures in the distribution function. Known examples in the solar wind include the scattering of heat-flux-carrying strahl electrons into the halo population [Table 1C; 223] or the reduction of the core-halo drift [224, 225]. In order to understand the physics of this kinetic multi-step heat-flux regulation, we must, therefore, measure the *fine structure of the distribution function* with an accuracy that allows us to identify small changes in the third velocity moment. Such measurements are only accessible through *in-situ* particle detectors. Statistical investigations of the electron heat flux, which also study its dependence on the plasma parameters, will allow us to distinguish the relevant heat-flux-regulation mechanisms. For example, we must understand the transitions from collisional to collisionless heat-flux regulation as well as the transition from heat-flux regulation through strahl-scattering to regulation through halo-scattering. It is also required to explore regimes in which the plasma reaches the free-streaming heat flux in order to make reliable predictions for the heat-flux value in other astrophysical objects. Moreover, we must investigate plasma regimes in which, for example, electron heating and expansion effectively increase the electron heat flux and counter-act the mechanisms that limit or reduce heat flux.

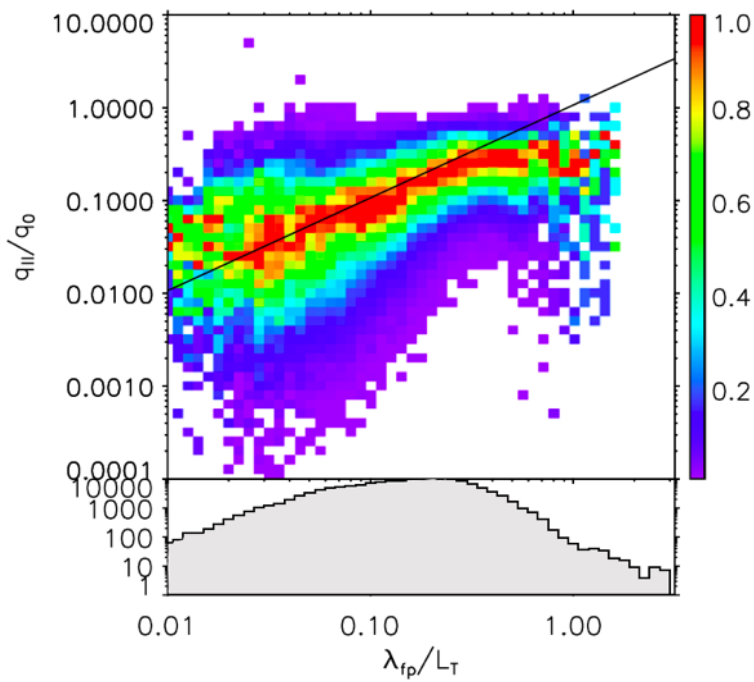

Figure 11: Field-parallel heat flux (normalised to the free-streaming value) as a function of normalised collisional mean free path. The straight line represents the Spitzer-Härm prediction, and the colour indicates the column-normalised probability found in solar-wind measurements. The heat flux deviates from the Sitzer-Härm prediction at large mean free paths. From Bale et al. [221].

If the amplitude of wave-like plasma fluctuations is large enough, electrons can be *trapped* in these structures and forced to bounce within the associated troughs in the electric potential [226-228]. Trapping suppresses the free streaming of the trapped electrons along the background magnetic field. Instead, these particles propagate with the speed of the electromagnetic structures they are trapped in; e.g., with the phase speed of the waves with respect to the background plasma. This speed can be substantially less than the thermal speed of the electrons. In this way, *trapping suppresses and controls electron transport*. Moreover, trapped electrons in turbulent structures and shocks can undergo efficient and localised *acceleration* to high energies [229, 230]. Therefore, we must investigate plasma intervals during which the wave amplitudes are large enough to trap a significant fraction of the electrons. High-resolution *in-situ* measurements of the electron distribution function in combination with detailed measurements of the trapping wave fields will then promote our understanding of the connections between electron trapping and heat-flux regulation in plasmas throughout the Universe.

### 2.5.4   Q4. What is the role of electrons in plasma structures and magnetic reconnection?

Magnetic reconnection is a fundamental plasma phenomenon occurring at thin plasma structures, called *current sheets*, in which magnetic-field energy is transferred to the plasma particles. Reconnection is important in plasmas throughout the Universe including the Sun [e.g., 231], solar wind [e.g., 232], magnetosphere [e.g., 233-235], and astrophysical plasmas [e.g., 236]. Reconnection requires both ions and electrons to decouple from the magnetic field and is, therefore, fundamentally a *kinetic electron-scale process* [234, 237, 238]. In the standard picture of reconnection in collisionless plasmas, particles fully decouple from the magnetic field in a region known as the *electron diffusion region* (EDR), which has a thickness comparable to $d_e$ (Figure 12). In a larger region with thickness $\sim d_p$ and encompassing the EDR, known as the *ion diffusion region*, the protons decouple from the magnetic field while the electrons remain frozen-in. While the EDR thickness is of order $d_e$, the size of the current sheet in the other dimensions is variable. Examples for large current sheets include those set up by large-scale interactions between solar wind and the magnetosphere [233]. Small-scale current sheets include those generated by turbulence [239-241].

Magnetic reconnection redistributes energy between thermal energy, bulk kinetic energy in the form of reconnection jets, and electromagnetic and electrostatic fluctuations. The partitioning of energy among these different channels and the particle species is an important open problem in the field of plasma astrophysics with major implications for electron-astrophysics. Observations of reconnection jets near Earth's magnetopause and in the magnetotail as well as numerical particle-in-cell simulations suggest that more thermal energy is imparted to ions than electrons. More specifically, these studies find that 13% of the available magnetic energy is converted to ion thermal energy and 1.7% to electron thermal energy [243-246]. It remains to be seen to what extent the same conclusion remains valid for reconnection under different circumstances; e.g., in solar-wind current sheets.

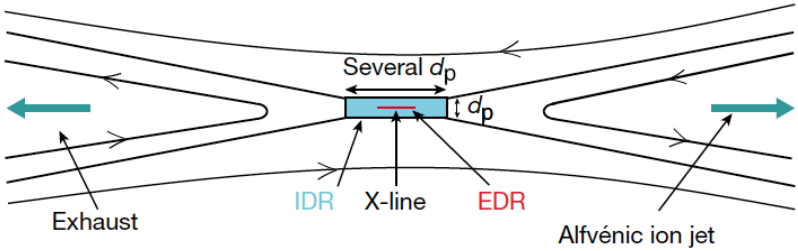

Figure 12: The standard picture of magnetic reconnection. The IDR is much larger than the EDR. However, if the current sheet is smaller than a few $d_p$, electron-only reconnection can occur. After Phan et al. [242].

Magnetic reconnection generates a variety of *secondary structures* that play an important role in electron energisation, including waves and turbulence in the outflows and along the separatrix [247-249], parallel electric fields and electrostatic structures [e.g., electron phase space holes, double layers, and solitons; 250-254], and Fermi acceleration in contracting magnetic islands [255]. These acceleration processes occur across many scales, extending down to the Debye length in the case of electrostatic structures. We must, therefore, obtain high-resolution measurements of the electron distribution to determine the details of the alterations in the fine-scale structure of the particle distributions and the plasma heating mediated by these processes. The particle acceleration facilitated by reconnection is significantly inhomogeneous [256] and also occurs far from the diffusion region itself due to the interaction of the reconnection jets with the surrounding environment [87], waves generated by the reconnection event, and processes occurring at the separatrices. This inhomogeneity leads to difficulties in the ability to quantify heating through reconnection and requires detailed observations throughout the reconnection outflows.

A novel form of *electron-only reconnection*, in which the ions do not interact with the reconnection dynamics, resulting in a lack of ion jets, was recently discovered in an interval of magnetosheath data observed by MMS [242]. Electron-only reconnection was observed at multiple thin current sheets ($\sim 4d_e$ in thickness) generated by magnetosheath turbulence. These events are only identifiable due to the high-resolution and multi-point electron measurements available from MMS, which allow the observation of thin, oppositely directed electron jets. The physics of electron-only reconnection is still unclear. One possible scenario suggests that electron-only reconnection occurs when the *length of the reconnecting current sheets* along the outflow direction is too short for ions to effectively couple to the reconnected field. Plasma simulations find that this effect sets in and results in weakened ion jets when the current sheet length $< 40d_p$, and ion jets are absent when the current sheet length $< 10d_p$ [257]. In a turbulent plasma, we approximate the length of the current sheets through the correlation length of the magnetic fluctuations. In the presence of the observed electron-only reconnection, this approximation is consistent with the $\sim 10d_p$ correlation length during the measurement interval [258]. The study of electron-only reconnection and the necessary conditions for its existence require more detailed examinations of the turbulent current sheets in the magnetosheath and other plasma environments such as the solar wind, in which the correlation length is much longer than in the magnetosheath. Moreover, the acceleration and heating associated with electron-only reconnection have yet to be quantified. We expect from the lack of ion interactions that any heating would be imparted largely to the electrons. Such a lack of ion heating would have significant implications for the *partition of energy between ions and electrons* in plasma turbulence. Taking the complex nature of the heating associated with reconnection as an indication, even higher-resolution measurements of the electron distributions than presently available will be necessary to explore electron heating in electron-only reconnection.

# 3 Potential space mission profiles

## 3.1 Mission requirements

In this section, we first discuss the general requirements for any mission to answer the science questions described above. We use Tables 2 and 3 to show the traceability from science questions to instrument performance. In order to address the 3D nature of waves and fluctuations as well as reconnection and heat conduction, it is necessary to make some of these measurements *simultaneously at multiple points in space, requiring multiple spacecraft*. This is the major factor that drives the complexity and cost of any mission in the field of electron-astrophysics. We differentiate between Small (S)-, Medium (M)-, and Large (L)-class missions that can be used to address the questions of electron-astrophysics in the final three subsections below.

Table 2: Linking the science objectives to measurement requirements to study electron-astrophysics.

| Science questions | Observational tasks | Measurement requirements | | | | | |
|---|---|---|---|---|---|---|---|
| | | R1 | R2 | R3 | R4 | R5 | R6 |
| Q1: What is the nature of waves and fluctuations at electron scales in astrophysical plasmas? | T1.1: Determine amplitudes, wavevectors, and frequencies of electromagnetic fluctuations. | S | S | S | | | |
| | T1.2: Determine amplitudes, wavelengths, and polarisations of electrostatic fluctuations. | | S | S | | | |
| Q2: How are electrons heated and accelerated in astrophysical plasmas? | T2.1: Identify signatures of electron-heating and acceleration processes. | S | S | S | | S | M |
| | T2.2: Measure partitioning of energy between ions and electrons. | S | S | S | S | S | |
| Q3: What processes determine electron heat conduction in astrophysical plasmas? | T3.1: Measure electron heat flux. | | | | S | S | M |
| | T3.2: Identify signatures of kinetic electron instabilities. | S | S | S | S | S | M |
| Q4: What is the role of electrons in plasma structures and magnetic reconnection? | T4.1: Observe small-scale current sheets and related structures. | M | M | M | L | L | |
| | T4.2: Measure electron dynamics. | | | | | L | M |

Analysis of our science objectives leads to the identification of eight specific *observational tasks* (T1.1-T4.2), which drive six specific *measurement requirements* (R1-R6). These links are summarised in Table 2 with S showing requirements for small missions, M for Medium missions, and L for Large missions.

Table 3: Measurement requirement traceability to the enabling instrumentation.

| Measurement requirements | Performance requirements | Enabling instrumentation |
|---|---|---|
| R1: Vector magnetic-field fluctuations | Frequency range: from 1.6 Hz to 3.2 kHz. Sensitivity: better than ($10^{-4}$, $10^{-6}$, $10^{-8}$, $10^{-10}$, $4\times10^{-11}$) nT$^2$/Hz at (1, 10, 100, 1000, 5500) Hz. | Search-coil magnetometer |
| R2: Vector electric-field fluctuations | Frequency range: from 0.01 Hz to 50 kHz. Sensitivity: better than ($10^{-11}$, $3\times10^{-14}$, $2\times10^{-14}$, $10^{-14}$, $8\times10^{-15}$) (V/m)$^2$/Hz at ($10^1$, $10^2$, $10^3$, $10^4$, $10^5$) Hz. | Electric-field probes |
| R3: Low-frequency and background magnetic field | Frequency range: from DC to 64 Hz. Sensitivity: better than $10^{-4}$ nT$^2$/Hz at 1 Hz. Accuracy of the background field: better than 0.1 nT in magnitude and 1º in direction. | Fluxgate magnetometer |
| R4: Proton moments and pressure tensor | Cadence: <1 s. Energy range: 200 eV to 4 keV. Energy resolution: $\Delta E/E < 0.1$. Angular resolution: $\leq 5.7º$. | Thermal-proton analyser |
| R5: Electron distribution functions and moments | Cadence: <1 ms pitch angle & moments, <10 ms 3D distribution. Energy range: 10 eV to 30 keV. Energy resolution: $\Delta E/E < 0.15$. Angular resolution: $\leq 11.3º$. | Thermal-electron analyser |
| R6: Energetic-electron distribution function | Cadence: 15 s. Energy range: 20 keV to 500 keV. Resolution: $\Delta E/E < 0.2$. Angular resolution: $\leq 45º$. | Energetic-electron analyser |

Each measurement requirement consists of a set of *necessary measurement characteristics* (e.g., cadence, sensitivity, resolution, and accuracy) that drive technical requirements on the payload complement and mission performance. We show the traceability from our measurement requirements to the instrument specifications and a summary of the performance requirements, discussed in the following subsections, in Table 3.

The key design principle for an electron-astrophysics mission is to sample plasma on timescales and length scales relevant to electron dynamics. Therefore, the requirements typically lead to shorter timescales and smaller distances than those sampled by previous missions. The performance requirements in Table 3 have been calculated assuming an electron scale of interest of order $l = 700$ m (e.g., $l \approx d_e$) and a solar-wind speed of $U = 700$ km/s. The sampling time required is defined as $t = l/U$, so that sampling of a structure of size $l$ in time and space requires a sampling time of just 1 ms with multiple measurement points inside an electron inertial length of each other. For electromagnetic plasma waves propagating across the spacecraft or between spacecraft, the requirements are more complex, but reduce to a similar argument, with sampling frequency required to distinguish a wave of approximately 2 kHz (a typical value for the electron gyro-frequency) over a propagation time of 1 kHz. Thus, a sampling frequency of 5 kHz is sufficient to provide a Nyquist frequency above 2 kHz. In order to measure the spatial structures and separate convected wavevector structures from frequency for propagating modes, *multi-spacecraft formation flying will be required*, with inter-spacecraft separations as small as possible, starting at only 300 m and extending to as large as 1000 km to sample electron effects associated with proton physics.

The enabling instrumentation indicated in Table 3 can be derived from a long heritage of existing space hardware. Search-coil and fluxgate magnetometers have flown on many missions, and only minor modifications would be required to meet the needs of an electron-astrophysics mission. The same can be said of electric-field probes and proton detectors. The payload that may require some further development are the *sufficiently fast and high-resolution electron analysers*. Such developments are discussed as part of Section 5 below.

### 3.2 Measurement environment, design, and orbits

In order to make results relevant to the wide range of physical systems discussed in Section 2, all missions designed to meet our goals must sample plasma with a wide range of $\beta$. Many space environments provide such a variation in $\beta$. However, in order to simplify the instrument design and to allow for a very direct physical interpretation of the physics results, avoiding dynamically complex regions like the Earth's magnetosheath or the ion and electron foreshock regions is ideal. Therefore, the *pristine solar wind* is the perfect plasma to study electron-astrophysics. Under these conditions, we recommend a wide Earth orbit or a deep-space orbit such as station-keeping at L1 or L2. L1 and L2 are ideal locations as a multi-spacecraft formation can maintain *very close proximity with reduced risk* and requires fewer orbital manoeuvres to maintain close proximity. We require that the key measurements of the high-frequency magnetic and electric fields as well as the electron distribution function be made in multiple locations. This will allow us to identify dynamic structures and waves and to search for dynamic features at electron scales. Since the spacecraft stay within a few hundred kilometres of each other for the entire duration of such a mission, the protons and the low-frequency magnetic field can be reliably measured at one location only. This less demanding requirement results from the fact that a proton gyro-radius is of order 100 km, and so spacecraft with a separation smaller than this distance will encounter very similar proton populations most of the time.

If the multi-spacecraft observatory stays in the solar wind for the majority of the mission lifetime, a *total mission duration of around 2 years* will be required to sample a wide range of solar-wind $\beta$. This constraint is derived from the distribution of proton $\beta$ measured over more than 20 years by the Wind spacecraft at 1 au. Since we expect more variation of the plasma parameters at 1 au during solar maximum, such a period would be slightly beneficial to maximise the coverage of plasma parameters.

The main challenge for electron-astrophysics missions in terms of their measurement environment is ensuring *electromagnetic cleanliness* for all spacecraft carrying the sensitive search-coil and fluxgate magnetometers and the electric-field probes. Electromagnetic emissions must be minimised during periods of data gathering. Electrons are easily deflected and accelerated by stray electric and magnetic fields, and so the *surface potential* of all spacecraft should be kept constant to within 1 V. Moreover, the potential with respect to space due to *spacecraft charging* by photoelectron emission and environmental interactions should be kept to a minimum. This does not necessarily require active control if the spacecraft is well designed for solar-wind studies, as for Solar Orbiter and for THOR in the ESA M4 mission Phase A study [259].

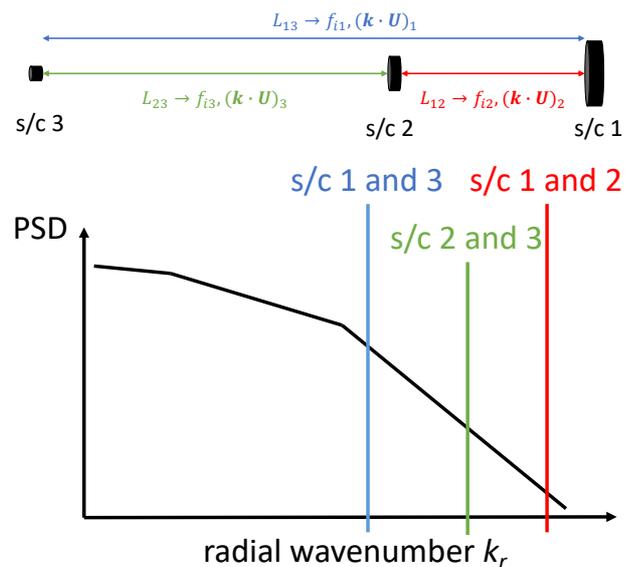

Figure 13: Simultaneous multi-scale measurement with three radially-aligned spacecraft. This setup assumes $L_{12} < L_{23} < L_{13}$, where $L_{ij}$ is the distance between spacecraft $i$ and $j$ as shown at the top.

## 3.3 Small-class mission (€150M)

The minimum practical electron-astrophysics mission consists of two spacecraft, separated along the spacecraft-Sun direction. A larger *main spacecraft (MSC)* carries a full science payload as required by the S symbols in Table 2. A single *deployable small-sat (DSS)* carries only a search-coil magnetometer. This configuration allows the identification of electromagnetic waves at electron scales via the high-frequency variation of magnetic field with a wavevector component along the direction of flow of the solar wind, as well as a full set of observations of particles and fields at the MSC. A similar mission design is discussed in detail in the Debye ESA F-class mission proposal [260].

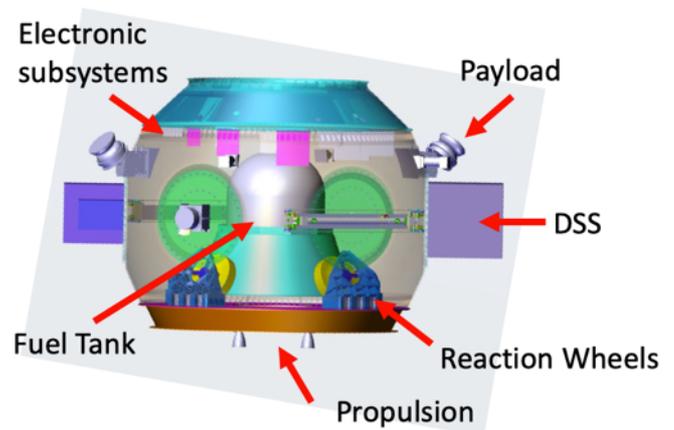

Figure 14: APMAS MSC design with payload and internal subsystems displayed.

**Table 4: S-class design key parameters.**

| Subsystem | Mass (kg) | Power (W) | TRL |
|---|---|---|---|
| Structure | 300 | - | 9 |
| Power | 45 | 45 | 9 |
| Data handling | 34 | 71 | 9 |
| Communication | 31 | 124 | 9 |
| Thermal | 31 | 10 | 6 |
| AOCS | 30 | 30 | 9 |
| Propulsion | 57 | - | 9 |
| Payload | 40 | 55 | 6 |
| Payload supports (e.g. boom) | 28 | - | 5 |
| Harness | 23 | 7 | 9 |
| **Total** | **619** | **342** | **-** |
| Margin (15%) | 93 | 51 | - |
| **System total** | **712** | **393** | **-** |
| DSS (x2) | 40 | | 6 |
| Propellant | 80 | | - |
| **Baseline wet mass** | **832** | | |

The DSS can be based on CubeSat hardware and have a mass of less than 20 kg. Adding a second DSS greatly increases the science return of the mission by allowing *multiple wavevectors to be distinguished simultaneously* (see Figure 13). Adding a third DSS increases the science return again by allowing a tetrahedral formation to *distinguish wavevectors in three dimensions*. Considering these factors, the best scientific return for a feasible mission within a Small-class mission budget consists of the MSC and two DSS. The DSS can be ejected from the MSC so that spacecraft separation can begin effectively at 0 m, allowing the smallest scales to be measured with unprecedented accuracy. A total launch mass of around 830 kg is achievable with margins (Table 4). Data from the DSS must be transmitted to the MSC for storage and eventual transmission to the ground.

The MSC itself could be based on existing Airbus APMAS architecture, effectively using the spacecraft dispenser from a primary launch as the spacecraft bus for a Small-class mission. Up to 4 DSS can be attached to the outer rim of the primary APMAS structure, with avionics, power, propulsion, communications, and data-handling systems located inside the hollow ring structure (Figure 14). Payload systems can be attached to the outside. An initial detailed analysis of this design has been made [260]. The key parameters are summarised in Table 4, which includes mass, power, and technological readiness of the various spacecraft subsystems. A mission based on this architecture could feasibly begin development immediately, and technological improvements (outlined below) will make such a mission design considerably simpler to implement in the time frame of Voyage 2050.

International participation by non-ESA nations or space agencies in a mission with multiple small-sats can be easily accommodated through the *modular structure* of the mission. Partners can build DSS units to meet requirements specified by the mission teams and ESA. This modularised approach reduces mission costs, for example by partnership with JAXA or NASA (see Section 5).

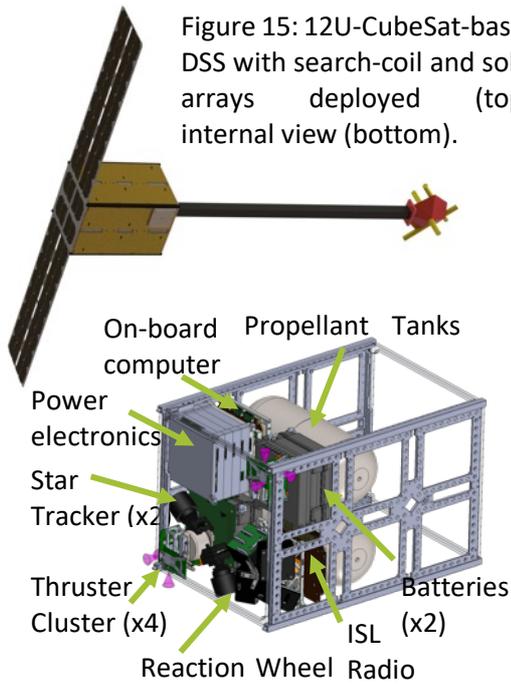

Figure 15: 12U-CubeSat-based DSS with search-coil and solar arrays deployed (top); internal view (bottom).

### 3.4 Medium-class mission (€550M)

An extended electron-astrophysics Medium-class mission design consists of an MSC and larger, *more capable DSS* carrying more extensive payload. In this case, an *energetic-electron instrument* will be added to the MSC to investigate electron acceleration to keV energies. Each of the DSS will carry electric-field probes and fluxgate magnetometers as well as search-coil magnetometers. Carrying a full suite of fields instruments on the DSS will allow us to investigate in more detail the three-dimensional structure of electrostatic fluctuations (Q1), current sheets (Q4), and electric fields that dissipate energy into electron heating (Q2). This enhanced instrument complement will also provide us with more detailed information on electron heat-flux instabilities and kinetic effects (Q3). These increased capabilities translate into an increased size and mass of the DSS. Through these design changes, the DSS mass will increase to between 50 and 70 kg, similar to small-sat designs such as the existing Surrey Satellites DoT IV bus or the JAXA Procyon mission. The design suggested here could transmit data from the DSS to the MSC for storage and later transmission to the ground, as in the case of the Small-class mission. However, the DSS still do not include the most demanding instrument, the thermal-electron analyser, which is only included on the MSC. The same MSC design as for the proposed Small-class design can be used, as the APMAS structure supports up to 300 kg of attached small-sats. Consequently, increasing the mass of the DSS does not require changes to the MSC design, apart from the higher telemetry demands of this configuration.

### 3.5 Large-class mission (€1200M)

The study of electron-astrophysics alone may not require a dedicated Large-class mission. However, it would greatly benefit from a Large-class mission consisting of *four (or more) identical spacecraft of the MSC design* or similar with electron-scale separations between some of the spacecraft. A mission of this type would allow the direct measurement of reconnection sites in 3D and the measurement of full wavevector information for electrostatic and electromagnetic fluctuations at small scales, with multi-point electron distribution measurements to investigate the 3D structure of electron populations. Such a mission would also target science goals described in other Voyage 2050 White Papers, for example, the *multi-scale coupling* and *energy transfer* in plasmas as well as the dynamics, heating, and acceleration of protons and electrons. Although the combination of these other concepts with our science goals may require an extension to the proposed payload, the science objectives described here could also be investigated with such a multi-spacecraft mission, provided the measurement requirements in Table 3 are met. With such an extended payload, other orbits could also be considered, for example a high Earth orbit that would facilitate measurements in the outer magnetosphere, magnetosheath, foreshock, and solar wind. The main barrier to such a mission is the very demanding telemetry data rate to downlink high-cadence 3D electron distributions with present-day communications technology. As described above, these measurements will produce a very large amount of data (>400 Gbit / day), which is a major challenge in terms of downlink time, especially when combined with data that may be necessary for the extended science goals of a combined Large-class mission.

## 4 Worldwide context

A dedicated mission for electron-astrophysics, making measurements in the near-Earth solar wind, has never been attempted before. The ESA Cluster mission is based on the concept of a *multi-spacecraft constellation to separate spatial and temporal features* of space plasma and has been a great success. The time resolution of the particle instruments on Cluster is not sufficient to identify small-scale features important for the science goals of electron-astrophysics. The NASA Magnetospheric Multiscale (MMS) mission is another multi-spacecraft mission. It is designed to observe reconnection in the Earth's magnetosphere. MMS instrumentation has far higher time resolution than Cluster, and the spacecraft orbit in a tighter formation than Cluster, with the specific purpose of observing the electron diffusion region in reconnecting current sheets. The mission has been very successful, with many high-impact publications in the last 5 years. However, the MMS payload is optimised for the magnetosphere and magnetosheath, regions which may be representative of some astrophysical objects, but not of the large-scale ambient plasma similar to the ICM or large objects such as accretion discs around compact objects. The disturbed plasma of the magnetosheath is far from equilibrium, making it difficult to extrapolate from the MMS data to answer our electron-astrophysics science questions. *Therefore, it is of prime importance to operate a multi-spacecraft electron-astrophysics mission in a space plasma like the solar wind, which is not affected by the Earth's magnetosphere.*

Space scientists across the world have recognised that multi-spacecraft plasma missions are the key to unlock the most important science questions in this field. In the U.S., for example, multiple white papers in response to the *Plasma 2020 Decadal Survey* express the need for multi-spacecraft constellations to disentangle spatial and temporal structures in plasma [261-263]. In this context, the European space community has the opportunity to take a world-leading role in the use of multi-spacecraft missions for studying electron-astrophysics. Following this recommendation, ESA would significantly enhance the synergies between the strong astrophysics and space-physics communities in Europe, which are often separated in other research programmes. Both communities would join forces and use data from electron-astrophysics missions to *advance our understanding of the Universe*. Therefore, a strong representation of electron-astrophysics in ESA's research portfolio would also unlock synergies beyond the mission involvement; e.g., through a significant increase in the organisation of joint astrophysics and space-physics conferences and joint publication activities.

It is important to note that, while our mission concepts will provide thoroughly new observations and advance our understanding of astrophysical plasma throughout the Universe, we recommend in the context of Voyage 2050 to instigate a Large-class opportunity for a *Grand European Heliospheric Observatory*. By combining our electron-astrophysics mission designs with one or more missions from the fields of solar, heliospheric, magnetospheric, and ionospheric physics, this combined observatory will not only address major challenges in electron-astrophysics but provide rapid scientific advances in a *holistic approach* to the otherwise disjunct science fields that underpin our European and worldwide *space-weather requirements* for decades to come. For example, the MSC design above would also work well as a space-weather monitor at L1, L2, or elsewhere at the end of its primary mission lifetime.

In order to ensure progress in electron-astrophysics on all fronts, we furthermore recommend to programmatically combine our *in-situ* electron-astrophysics missions with missions targeting electron physics signatures in astrophysical environments. These missions include *x-ray telescopes* for studying reconnection jets in AGNs, *space VLBI missions* to study electron-synchrotron radiation in pulsars and elsewhere, or *optical/polarisation missions* that image jets and determine the effects of strong magnetic fields on plasma. Combining such missions will help to bring the astrophysics and space plasma physics communities together to increase cross-fertilisation between these two major research fields.

# 5 Technology challenges

The measurement of high-cadence and high-resolution electron distribution functions as well as the ability to perform synchronised multi-spacecraft measurements are the main challenges for the mission designs described above. These measurements present several key problems: (i) the *sensitivity* of the detector to count low numbers of electrons in short times accurately, (ii) the *saturation* of the detector due to high count rates per second when detectors run at such high cadence, and (iii) the very large amount of *data* created. Other technology challenges that could help the mission operate more effectively are small-sat technologies, such as miniaturised systems, autonomous operations, data relaying, precision flying in deep space, and increased standardisation of the spacecraft-integration process. We describe the required work briefly below.

## 5.1 Scientific instrumentation

The primary detector for an electron-astrophysics mission is the thermal-electron detector. Thermal electrons arrive at the spacecraft from all directions, being a roughly isotropic population. In order to measure this diffuse population with efficient use of resources, detectors that sample 180º or 360º slices of the sky are typically mounted on spinning spacecraft, or in recent designs (EAS on Solar Orbiter, FPI DES on MMS), use electric fields at the aperture to *deflect* electrons and to scan across the sky. As the acceptance direction changes, the instrument uses a voltage to select electrons of different energies. Thus, the counts measured ($C$) at a specific energy ($E$), in a specific direction ($\theta, \phi$), depend on the accumulation time ($\delta t$), which is a function of the acceptance bin widths ($\delta\theta, \delta\phi$) and bin widths of the energy steps ($\delta E$).

*Spacecraft spin* cannot be used to measure the electron population at 1 ms cadence as the spacecraft cannot spin so rapidly. Current designs of electrostatic deflection systems require four or eight deflection steps to sample the entire sky. In this case, the accumulation time must divide 1 ms by 4 or 8 and then by the number of energy steps, typically a minimum of 32. Thus, the accumulation time for an energy step in an electrostatic deflection detector is roughly 4 $\mu s$. This number is problematic since the detector surface, for example a *microchannel plate (MCP)* or *channel electron multipliers (CEMs)*, saturates at around $10^7$ counts per second. Thus, counting a statistically satisfactory number of electrons (say 100) in the peak will inevitably lead to saturation for a traditional design of instrument. *Instrument development* is required to find new methods to reduce the count rate, without reducing the total number of counts or the time cadence of the full distributions. There are three ways to approach this problem:

**Improve detector technology to increase the saturation count rate.** Work is required to increase the efficiency of MCP and CEM detectors, and the supporting anodes and electronics (as has been done at LPP in France, for example, with the use of ASICS in the detector-anode systems) to increase the saturation count rate.

**More detector units and acceptance directions operating simultaneously.** If more angular directions are sampled simultaneously, the *integration time per direction* can increase and so the overall count rate decreases. Solar Orbiter EAS uses two detector heads and eight deflection states. MMS FPI DES uses four heads with four deflection states each. The next generation of instruments will need *more distributed heads* (high demands in terms of cost, mass, and power), or to accommodate *more separate look-directions* in a single unit. Testing work on such a method is underway in France but needs to have TRL raised to be used for an ESA mission.

**Sample energies simultaneously.** Most of the time steps in the electron-distribution measurement are required to cover the energy range using the electrostatic analyser concept. Detectors that use magnetic fields or more novel designs of electrostatic aperture have been proposed that could sample some or all of the electron energy spectrum simultaneously, thus reducing the integration time by up to a factor of 32 over current designs. These designs are promising for an electron-astrophysics mission but are currently at low TRL, typically 2 or 3. A specific effort to develop designs based on these systems to flight readiness is recommended.

## 5.2 Spacecraft bus

**Increase spacecraft data-downlink rate.** Because electron-astrophysics occurs at short timescales and small spatial scales, the science data is necessarily recorded at much higher frequency than in classical space-plasma missions. Therefore, even our Small-class concept will generate around 500 Gbits of data per day. This amount of data is currently possible to downlink from deep space, but only with a large high-gain antenna on the spacecraft and using 35 m-class ground stations. Increasing the capacity of ESA and other international facilities to downlink data from spacecraft at large distances at high rates would be very beneficial for this field and many others. Recent developments in *optical data transmission* show great promise at increasing the data rate to levels high enough to easily accommodate our mission designs. Any increase in the efficiency or decrease in the cost of deep-space communications would inherently make these missions more feasible.

**Command and data relay for small-sats in deep space.** The use of small-sats in deep space requires new methods of communication as well as command and control. It is unlikely that the small-sats will be able to communicate directly with the ground effectively enough for science-data transmission. Thus, developing more advanced *data-relay capability* for the MSC satellite bus may be required. Automated or semi-automated operational procedures for science and command and control of multiple small-sats in space would reduce the workload for the ground-control teams and make multi-spacecraft missions more feasible.

**Autonomy and coordination of small-sats.** Keeping multiple small-sats in close proximity in deep space requires a level of *autonomous operations,* specifically regarding range-finding, navigation, and collision avoidance. The small-sats must also be able to coordinate time keeping and location data for accurate science data.

**Spacecraft integration.** Medium-to-Large-class electron-astrophysics missions require a large number of instruments to achieve the necessary simultaneous measurements of particles and fields. Significant spacecraft-integration loads are cost drivers and time consuming in the mission build phase. We recommend the development of more *standardised instrument interfaces* and *standardised instrument packages* to simplify and accelerate this process.

# A-1. Bibliography